\documentclass[letterpaper, comsoc, 10pt]{IEEEtran}

\usepackage{ifpdf}
\usepackage{cite}
\usepackage{subfigure}
\ifCLASSINFOpdf
	\usepackage[pdftex]{graphicx}
\else
  	\usepackage[dvipdfmx]{graphicx, color}
\fi
\usepackage[cmex10]{amsmath}
\usepackage[caption=false,font=footnotesize]{subfig}

\ifCLASSOPTIONcaptionsoff
	\usepackage[nomarkers]{endfloat}
	\let\MYoriglatexcaption\caption
	\renewcommand{\caption}[2][\relax]{\MYoriglatexcaption[#2]{#2}}
\fi
\usepackage{url}
\usepackage{amssymb}
\usepackage{amsthm}
\usepackage{mathtools}
\usepackage{booktabs}
\usepackage{color}
\usepackage{bm,bbm}
\usepackage{algorithm,algpseudocode}
\usepackage{ulem}

\usepackage{optidef}

\algdef{SE}[DOWHILE]{Do}{DoWhile}{\algorithmicdo}[1]{\algorithmicwhile\ #1}
\algdef{SE}[SUBALG]{Indent}{EndIndent}{}{\algorithmicend\ }%
\algtext*{Indent}
\algtext*{EndIndent}

\usepackage{comment}

\newcommand{\yama}[1]{
	}

\theoremstyle{definition}

\title{Communication-Efficient Multimodal Split Learning\\ for mmWave Received Power Prediction}
\author{
	Yusuke~Koda,~\IEEEmembership{Student~Member,~IEEE,}
	Jihong Park,~\IEEEmembership{Member,~IEEE,}
	Mehdi~Bennis,~\IEEEmembership{Senior Member,~IEEE}
	Takayuki~Nishio,~\IEEEmembership{Member,~IEEE,}
	Koji~Yamamoto,~\IEEEmembership{Member,~IEEE,}
	Masahiro~Morikura,~\IEEEmembership{Member,~IEEE,}
	Kota~Nakashima,~\IEEEmembership{Student~Member,~IEEE,}\vspace{-2.2em}
	\thanks{
		Y. Koda, K. Yamamoto, T. Nishio, M. Morikura, and K. Nakashima are with the Graduate School of Informatics, Kyoto University, Yoshida-honmachi, Sakyo-ku, Kyoto 606-8501, Japan (e-mail: \{koda@imc.cce., kyamamot@, nishio@, nakashima@imc.cce.\}i.kyoto-u.ac.jp).

		J. Park was with the University of Oulu, and now is with the School of Information Technology, Deakin University, Geelong, VIC 3220, Australia (email: jihong.park@deakin.edu.au)
		
		M. Bennis is with the Centre for Wireless Communications, University of Oulu,   90014 Oulu, Finland and is also with the Department of Computer Science (e-mail: mehdi.bennis@oulu.fi).

		This work was supported in part by JSPS KAKENHI (Grant No. JP17H03266, JP18H01442), and KDDI Foundation.
		This work was also supported in part by the Academy of Finland under Grant 294128, in part by the 6Genesis Flagship under Grant 318927, in part by the KvantumInstitute Strategic Project (SAFARI), in part by the Academy of Finland through the MISSION Project under Grant 319759, and in part by the NOKIA grant foundation.
	}
}

\begin{document}

\maketitle
\begin{abstract}
The goal of this study is to improve the accuracy of millimeter wave received power prediction by utilizing camera images and radio frequency (RF) signals, while gathering image inputs in a communication-efficient and privacy-preserving manner. 
To this end, we propose a distributed multimodal machine learning (ML) framework, coined \emph{multimodal split learning (MultSL)}, in which a large  neural network (NN) is split into two wirelessly connected segments. 
The upper segment combines images and received powers for future received power prediction, whereas the lower segment extracts features from camera images and compresses its output to reduce communication costs and privacy leakage. Experimental evaluation corroborates that MultSL achieves higher accuracy than the baselines utilizing either images or RF signals. 
Remarkably, without compromising accuracy, compressing the lower segment output by 16x yields 16x lower communication latency and 2.8\% less privacy leakage compared to the case without compression.
\end{abstract}


\begin{IEEEkeywords}
	Millimeter-wave communications, received power prediction, multi-modal deep learning, split learning.
\end{IEEEkeywords}

\vspace{-1em}
\section{Introduction} \label{sec:introduction}
\IEEEPARstart{W}{ireless} communication systems can benefit from peripheral data source information in addition to the radio frequency (RF) signal domain, such as location, motion sensory data, and camera images \cite{huang2013handover,lehne2016analyzing,ei2010trajectory, taha2017intelligent,nishio_jsac}. Incorporating these non-RF modalities can complement insufficient features in RF signals, enabling more accurate handover decisions \cite{ei2010trajectory}, received power predictions~\cite{nishio_jsac}, and so on. 
In view of this, in this letter we focus on the problem of millimeter-wave (mmWave) uplink received power prediction by efficiently integrating the received mmWave RF signal powers and depth camera images. 

\begin{figure}[t]
	\centering
	\subfigure[Baseline 1: \textsf{RF}.]{\includegraphics[width=0.21\textwidth]{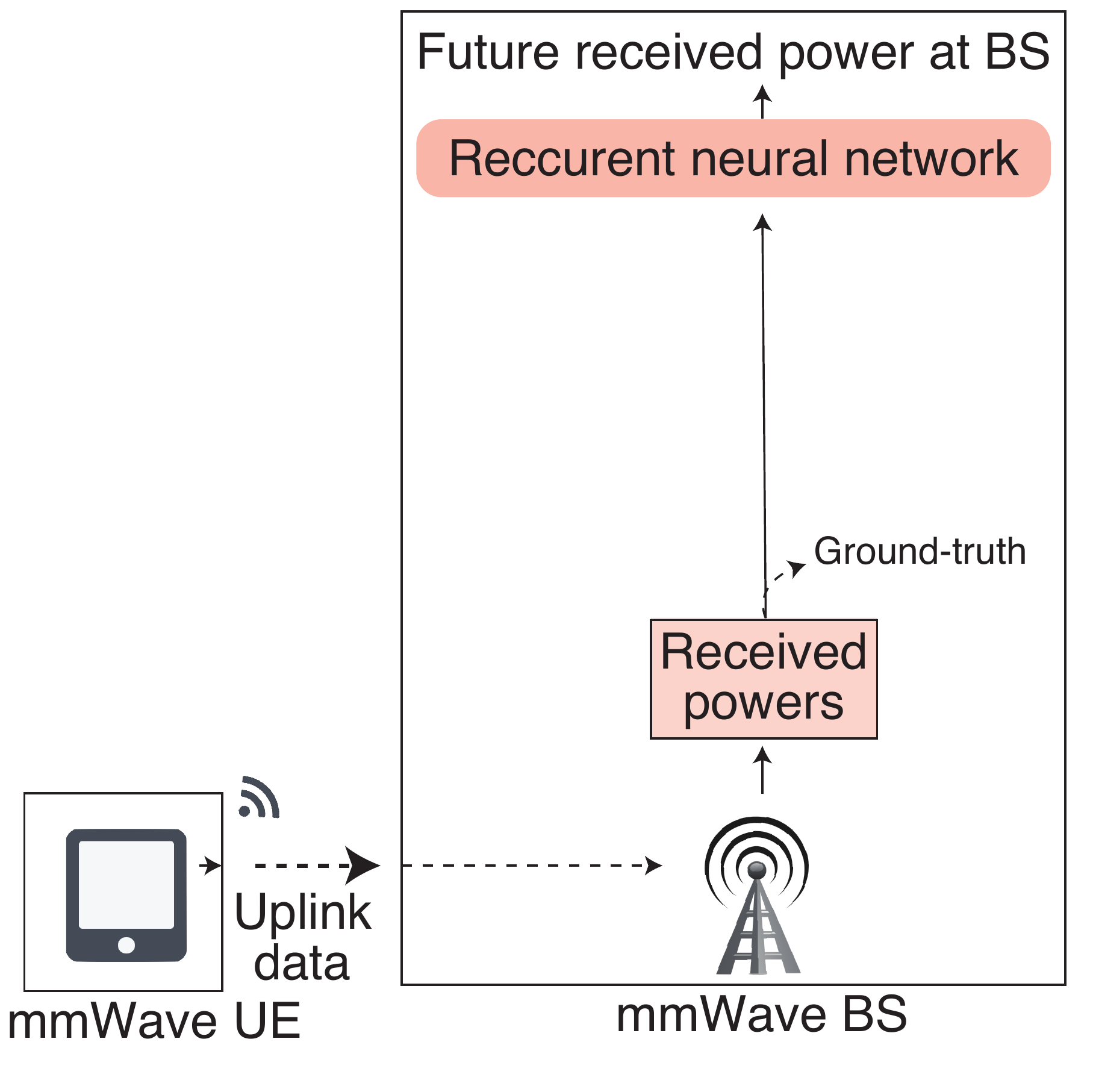}}
	\subfigure[Baseline 2: \textsf{Img}.]{\includegraphics[width=0.203\textwidth]{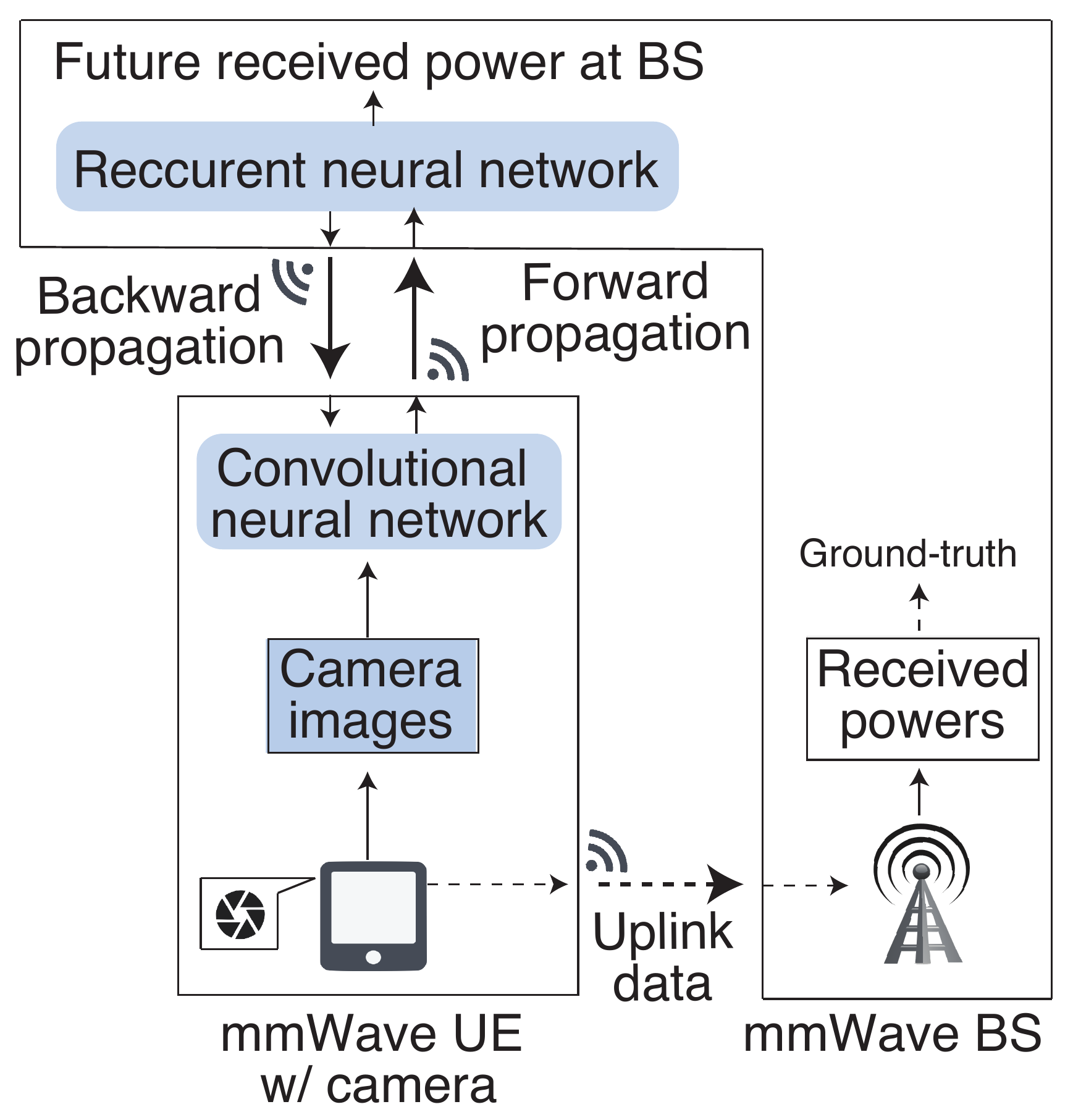}}\\
	\subfigure[Proposed: \textsf{Img+RF}.]{\includegraphics[width=0.43\textwidth]{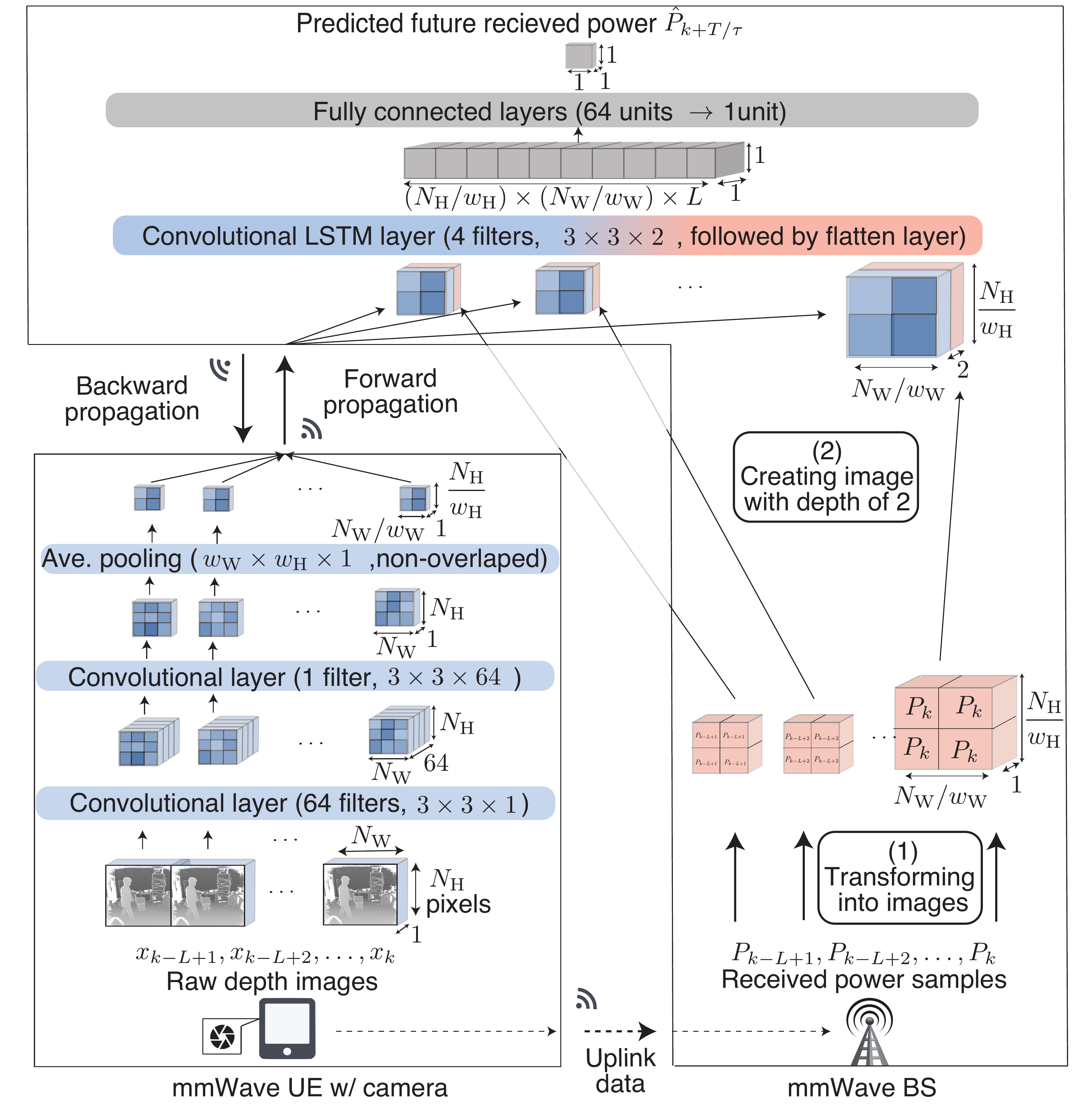}}
	\caption{
		Neural network (NN) architectures for mmWave power prediction: based solely on (a) previously received RF signal powers (\textsf{RF}) and on (b) depth images (\textsf{Img}), compared to (c) our proposed multimodal split learning (MultSL) based on both RF signal powers and depth images (\textsf{Img+RF}).
		Convolutional layers in UE process depth images one by one and thereby extract spatial features per image, which are separated from the recurrent layers termed ``convolutional LSTM layer''.
		Image feature maps extracted by UE are integrated with the received power values as follows: 1) a received power is transformed into an image filled by its value whose shape is the same as that of a depth image feature map from UE. 2) An image with the depth of two is created wherein the first and second depth correspond to the received power and depth image feature map from UE, respectively.
	}
	\label{fig:system_model}
	\vspace{-2em}
\end{figure}

\begin{figure*}
	\centering
	\subfigure[Raw images.]{\includegraphics[width=0.23\textwidth]{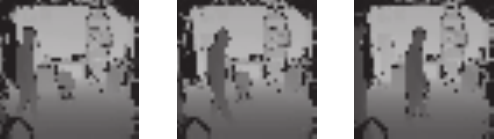}}\hspace{8pt}
	\subfigure[$w_\text{H}\!\times\! w_\text{W}$: $4\!\times\! 4$.]{\includegraphics[width=0.23\textwidth]{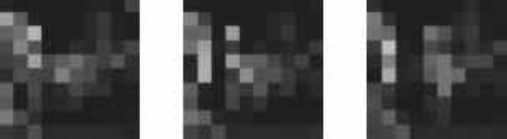}}\hspace{8pt}
	\subfigure[$10\!\times\! 10$.]{\includegraphics[width=0.23\textwidth]{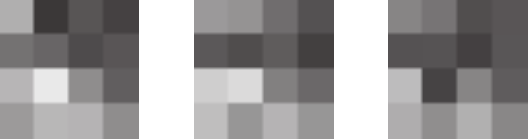}}\hspace{8pt}
	\subfigure[$40\!\times\! 40$.]{\includegraphics[width=0.23\textwidth]{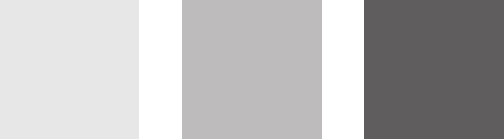}}
	  \caption{Raw depth-images and output images of trained convolutional layers in MultSLs with different pooling dimension $w_{\mathrm{H}}\times w_{\mathrm{W}}$.
	  Trained convolutional layers extract the image feature representations according to the given pooling dimensions.}
	  \label{fig:images}
	  \vspace{-1.5em}
	\end{figure*}

As shown by a prior work\cite{nishio_jsac}, depth image-based prediction exploiting machine learning (ML) reaches better accuracy by recognizing mobility blockage patterns to detect sudden changes between line-of-sight (LoS) and non-LoS conditions, which is hardly observable from received mmWave signal powers. 
By contrast, current received mmWave signal powers are useful for predicting short-term received power fluctuations for a given LoS or NLoS condition~\cite{11ad_channel}. To reach their full potential, our goal is to fuse both RF received powers and depth images in an ML-based received power prediction. 

There are two key challenges in acquiring depth images: communication latency and privacy violation.
	The first challenge is due to the fact that depth images are not necessarily obtained in the same location of the RF received power.
	The physical separation necessitates communication between the entity holding the images (e.g., user equipment (UE) or surveillance cameras) and that holding RF received powers (e.g., base stations (BSs)) over a limited wireless bandwidth, and this can cause a severe latency in the collection of depth images.
	However, numerous applications for mmWave communications are delay-sensitive (e.g., virtual reality\cite{wang2018millimeter}).
	Hence, it is important to design a prediction framework with lower communication latency for acquiring depth images.
	The second challenge is due to the fact that depth images may also involve privacy-sensitive information, e.g., the travel history of people in the view of cameras. 
	Therefore, acquiring raw depth images violates the privacy of the pedestrians who block mmWave links, which motivates us to design a framework to perform received power prediction in a privacy-preserving manner.

To address the aforementioned challenges, we propose a communication-efficient and privacy-preserving \emph{multimodal split learning (MultSL)} framework. Exploiting a split NN architecture~\cite{gupta2018distributed}, without sharing raw data, MultSL combines RF and image modalities by only exchanging NN activations and gradients (Fig.~\ref{fig:system_model}(c)). 
Before exchanging NN activations, the last activations for the image modality are compressed (see Fig.~2), achieving higher communication efficiency while preserving more data privacy. 
Surprisingly, experimental evaluations show that the compression is beneficial for balancing the fusion between RF and image modalities. Consequently, the MultSL with an optimal compression rate achieves higher accuracy, compared not only to baseline schemes based solely on either received mmWave powers (\textsf{RF}, Fig.~1(a)) or images (\textsf{Img}, Fig.~1(b)), but also to the MultSL without compression.

\vspace{3pt}\noindent\textbf{Related Works.} 
For handover or positioning, RF-based received power or channel state informaiton are utilized\cite{mmwave_mdp, kaltiokallio2017three}. 
For mmWave received power prediction, the prior study in \cite{nishio_jsac} utilizes camera images. 
While the aforementioned works consider a single modality, the proposed MultSL utilizes both image and RF modalities for mmWave received power prediction, thereby achieving higher accuracy.
Moreover, while the study in \cite{nishio_jsac} does not take into account a communication efficiency and privacy in gathering images, MultSL integrates image and RF modalities in a communication-efficient and privacy-preserving manner by leveraging a novel split learning (SL) framework.

The original SL framework in \cite{gupta2018distributed} combines NN activations and gradients that are generated from a single modality without exchanging raw data.
In \cite{vepakomma2019reducing}, to improve privacy guarantees in SL, the split NN is optimized to maximize the KL divergence between the distributions of raw health data and NN activations.
The aforementioned works focus on a single modality and do not consider communication efficiency.
In contrast, MultSL integrates NN activations originated from \textit{two different modalities} and optimizes the split NN by compressing the last activations of depth images; thereby improving communication efficiency and privacy guarantees.

\section{MultSL for Future Received Power Prediction}
\label{sec:system_model}
The MultSL structure is illustrated in Fig.~\ref{fig:system_model}(c), in which a convolutional long short-term memory (LSTM) NN is split and distributed across a UE and its associated BS. 
In Fig.~\ref{fig:system_model}(c), $k\in\mathbb{N}$ denotes the time index, $x_k$ denotes the observed image, $P_k$ denotes the corresponding received power in the uplink signal, and $\tau$ denotes the time-interval between successive images.
The UE is equipped with a depth camera whose captured images  $x_{k - L + 1}, x_{k - L + 2},\dots, x_k$ are processed by convolutional layers and then uploaded to the BS, where $L$ denotes the length of an image sequence. 
The BS stores an LSTM layer where the uploaded images are integrated with the uplink mmWave RF signal powers $P_{k - L + 1}, P_{k - L + 2},\dots, P_k$ received by the BS. 
By processing the integrated image and RF data at the last fully connected layer, the BS can predict its future mmWave received power $P_{k + T/\tau}$ with a look-ahead time $T$.

For training MultSL, the UE sends the output of its convolutional layers, termed a forward propagation (FP) signal to the BS.
Based on the FP signal, the BS subsequently calculates the gradients of the weight parameters in its own NN and sends the gradients, termed a backward propagation (BP) signal, back to the UE.
Finally, based on the BP signal, the UE updates the weight parameters of the convolutional layers.

The communication of the FP/BP signals is performed over a wireless channel. 
In this letter, we consider that the FP/BP signals are transmitted via the mmWave channel over which the aforementioned uplink signals are transmitted to achieve smaller transmission delay through exploiting a wider bandwidth in the mmWave channel.
Section~\ref{sec:experimental_results} further details the mmWave communication channel.

MultSL is compared with two baselines based solely on a single-modality as shown in Figs.~\ref{fig:system_model}(a) and \ref{fig:system_model}(b).
The former baseline termed as \textsf{RF} is based only on consecutive received powers in which the LSTM layer and fully connected layer are located at the BS, and training is performed at the BS.
The latter baseline, termed as \textsf{Img} is based only on consecutive images, in which the entire NN is split similarly in MultSL.

\vspace{-0.5em}
\section{Compression of CNN Output via Pooling Towards Communication Efficiency and \\Privacy Preservation}
\label{sec:communication_efficient_structure}

In addition to prediction accuracy, MultSL involves the following two key metrics: over-the-air latency for transmitting FP signals and data privacy\cite{gupta2018distributed}.
Hence, it is important to optimize the operation of MultSL such that both latency and privacy leakage are minimized without compromising the prediction accuracy.

We notice that increasing the pooling dimension $w_{\mathrm{W}}\times w_{\mathrm{H}}$ in the convolutional layer: i.e., the compression intensity of the convolutional layer output, is suitable for reducing both the latency and privacy leakage.
Given that the pooling dimension is $w_{\mathrm{W}}\times w_{\mathrm{H}}$, and pooling regions are non-overlapping (i.e., the horizontal and vertical strides are $N_{\mathrm{W}}/w_{\mathrm{W}}$ and $N_{\mathrm{H}}/w_{\mathrm{H}}$, respectively), the number of pixels to be forwarded to the BS is $LN_{\mathrm{W}}N_{\mathrm{H}} / w_{\mathrm{W}}w_{\mathrm{H}}$ as shown in Fig.~\ref{fig:system_model}(c).
Thus, the payload size of FP signals and the consequent over-the-air latency for transmitting them can be reduced by increasing the pooling dimension.
Moreover, the privacy leakage may be reduced by the increase of pooling dimension since compressed images make it harder to see what the images reflect (as depicted in Fig.~\ref{fig:images}).
In the next section, we experimentally demonstrate that an increasing pooling dimension yields a smaller latency and privacy leakage.

\vspace{-0.5em}
\section{Experimental Evaluation}
\label{sec:experimental_results}

\subsection{Setting}
\begin{figure}[t]
	\centering
	\includegraphics[width=0.65\columnwidth]{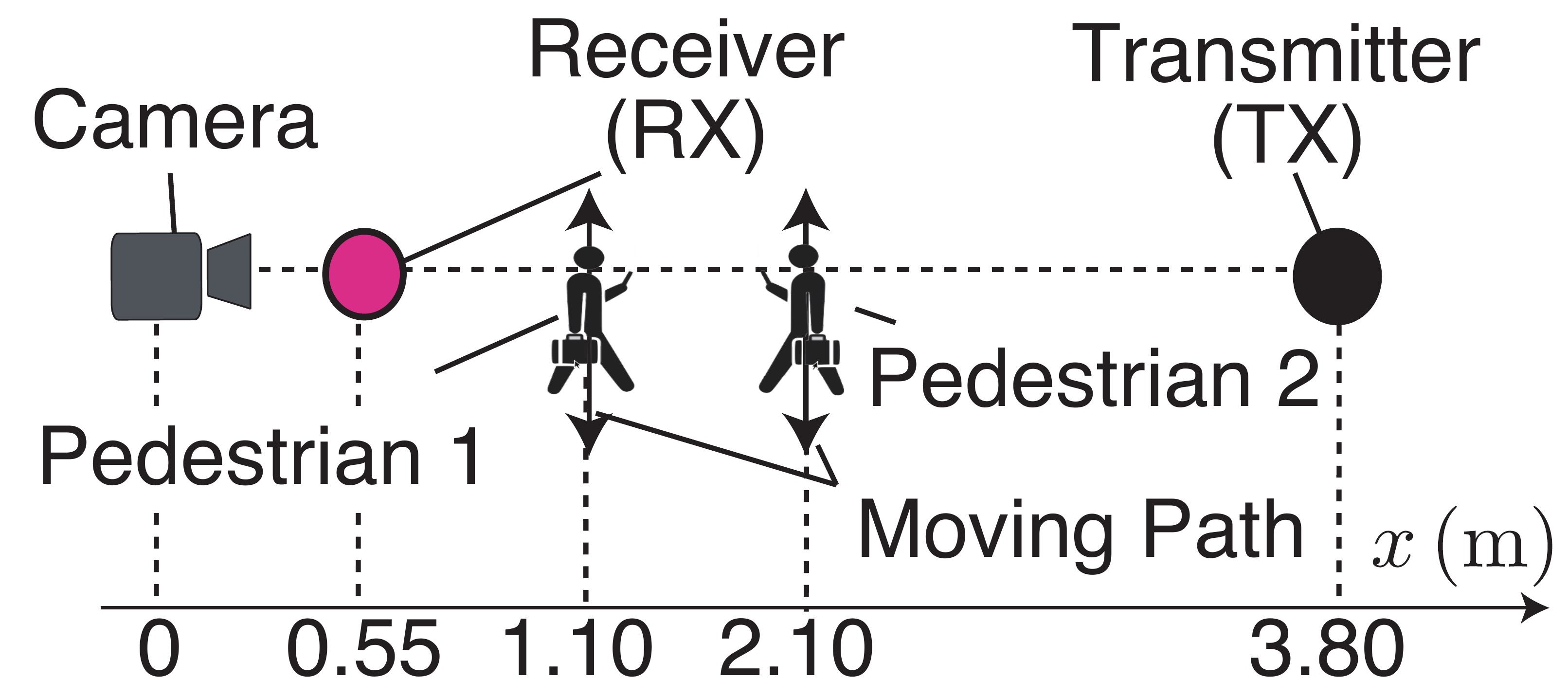}
	\caption{Experimental environment for measuring the communication channel and depth image data, in which a pair of mmWave transmitter and receiver is blocked by two moving pedestrians.}
	\label{fig:measurement_environment}
	\vspace{-1.5em}
\end{figure}

\noindent\textbf{Datasets.}
The prediction accuracy achieved by the proposed MultSL is evaluated using the data set of received powers and depth images\cite{nishio_jsac}.
The experimental environment is shown in Fig.~\ref{fig:measurement_environment}.
We deployed a transmitter (TX), receiver (RX), and camera.
As the TX, we utilized a commercial product of IEEE 802.11ad access points.
As the RX, we utilized the measurement system developed in \cite{koda_measurement2}.
The usage of this measurement system is because it is validated to be capable of measuring the time-variance of the mmWave received powers due to moving obstacles, which is essential to train the NN models in Fig.~\ref{fig:system_model}\cite{nishio_jsac}.
The TX transmits signals at the carrier frequency of 60.48\,GHz towards the RX, and two pedestrians walk across the path between the TX and RX.
The camera (Kinect sensor\cite{kinect2}) obtains depth images viewing the two pedestrians with the resolution of $512\times 480$ and with the interval of successive image frames $\tau$ of 33.3\,ms.
The obtained images are compressed to have a pixel resolution of $N_{\mathrm{W}}\times N_{\mathrm{H}}=40\times 40$.
The details of the measurement are discussed in \cite{nishio_jsac}.

\vspace{3pt}\noindent \textbf{Training, Validation, and Test.}
The training, validation, and test are performed with datasets that differ from one another.
It is noted that the validation is performed \emph{during training} to monitor the prediction accuracy while preventing overfitting. 
The test is performed \emph{after training} to evaluate the final performance of the trained model.
The procedures are both commonly performed in ML\cite{goodfellow}.
Let $k\in\mathcal{K}$ denote the time-index of the image and  received power samples obtained in the aforementioned measurement with $\mathcal{K}=\{1, 2,\dots, 15325\}$.
We perform training, validation, and test with samples whose time-index is in the index set $\mathcal{K}_{\mathrm{train}}$, $\mathcal{K}_{\mathrm{valid}}$, and $\mathcal{K}_{\mathrm{test}}$, respectively, wherein $\mathcal{K}_{\mathrm{train}}\cup \mathcal{K}_{\mathrm{valid}} \cup \mathcal{K}_{\mathrm{test}} = \mathcal{K}$.
In this evaluation, the ratio of $|\mathcal{K}_{\mathrm{train}}|$ and $|\mathcal{K}_{\mathrm{valid}}|$ is set as 75\% and 25\%, respectively, and that of $|\mathcal{K}_{\mathrm{train}}|$ and $|\mathcal{K}_{\mathrm{test}}|$ is set as 80\% and 20\%, respectively; hence, $\mathcal{K}_{\mathrm{train}} = \{1, 2, \dots, 9928\}$, $\mathcal{K}_{\mathrm{valid}}= \{9929, 9930, \dots, 13228\}$, and $\mathcal{K}_{\mathrm{test}} = \{9929, 9930, \dots, 15325\}$.

The training is performed such that the mean square error (MSE) between the predicted and actual received powers is minimal.
Let the actual received power at the time-index $k$ be denoted by $P_{k}$.
Given pooling dimension $w_{\mathrm{W}}\times w_{\mathrm{H}}$, we solve the following optimization problem:\vspace{-1em}
	\begin{mini}
		{\substack{\theta}}
		{\frac{1}{|\mathcal{K}_{\mathrm{train}}|}\sum_{k\in \mathcal{K}_{\mathrm{train}}}\bigl(\hat{P}^{(\theta, w_{\mathrm{W}}, w_{\mathrm{H}})}_{k + T/\tau} - P_{k + T/\tau}\bigr)^2}{}{},
	\end{mini}
	where $\hat{P}^{(\theta, w_{\mathrm{W}}, w_{\mathrm{H}})}_{k + T/\tau}$ denotes the predicted received power values with the look ahead time $T = 120$\,ms given the trained NN parameters $\theta$ and pooling dimension $w_{\mathrm{W}}\times w_{\mathrm{H}}$.
	Note that $\hat{P}^{(\theta, w_{\mathrm{W}}, w_{\mathrm{H}})}_{k + T/\tau}$ is calculated from the input of sequential images $x_{k - L + 1}, x_{k - L + 2}, \dots, x_k$ and received powers $P_{k - L + 1}, P_{k - L + 2}, \dots, P_k$ with the sequence length of $L = 4$.
	The problem is solved with the Adam optimizer\cite{sutskever2013importance} with the learning rate of $1.0\times 10^{-3}$, the decaying rate parameters $\beta_1=0.9$ and $\beta_2=0.999$, and the batch size of 64.
The training is continued until 50 training epochs (156 stochastic gradient descent steps) are iterated.
	In the MultSL and the baseline frameworks that are subsequently discussed, both UE and BS trained their NN layers, i.e., performed forward and backward calculations in parallel computing via exploiting an Nvidia Tesla T4 GPU with 2560 cores with memory corresponding to 16\,GB and memory bandwidth corresponding to 320\,GB/s.

\vspace{3pt}\noindent\textbf{FP/BP Communication Channel.} 
FP/BP signals are assumed to be transmitted over the mmWave channel. 
Under this channel, an NLoS event decreases the received power by around 15\,dB compared to the values in LoS conditions, and lasts for 200-300\,ms. The LoS-NLoS transition occurs within 50--200\,ms. An exemplary snapshot capturing these channel characteristics is illustrated in Fig.~\ref{fig:timeseries_w4}.


\vspace{3pt}\noindent\textbf{Baselines.} 
The proposed prediction framework is compared with two baselines that depend solely on either image sequences or received power sequences.
In the former baseline, termed as \textsf{Img}, only the output of the convolutional layers are fed into the convolutional LSTM layer.
In the latter baseline, termed as \textsf{RF}, only the received power values are fed into the convolutional LSTM layer.
We denote the proposed multimodal prediction framework as \textsf{Img+RF} to explicitly indicate that the proposed framework utilizes both images and received power values for the prediction.

It is noted that the objective of the comparison is to demonstrate the improvement in accuracy from using the other modality \textit{in addition to} a single modality.
Hence, although there are differences in input data sizes, \textsf{Img+RF} uses 8 inputs (image and received power sequences wherein each exhibits a length of 4) while \textsf{Img} and \textsf{RF} use 4 inputs (image or received power sequence with a length of 4).

\vspace{-1em}
\subsection{Performance metrics}
\noindent\textbf{Prediction Accuracy.} 
Prediction accuracies in validation and test are evaluated using the RMSE.
Given the predicted received powers $\Bigl(\hat{P}^{(\theta, w_{\mathrm{W}}, w_{\mathrm{H}})}_{k + T/\tau}\Bigr)_{k\in\mathcal{K}_{j}}$ in the trained parameters $\theta$ and pooling dimension $w_{\mathrm{W}}\times w_{\mathrm{H}}$ where $j\in\{\mathrm{valid}, \mathrm{test}\}$, the RMSE is given as follows:
\begin{align}
	\text{RMSE} = \sqrt{\frac{\sum_{k\in\mathcal{K}_{j}}\Bigl(\hat{P}^{(\theta, w_{\mathrm{W}}, w_{\mathrm{H}})}_{k + T/\tau} - P_{k + T/\tau}\Bigr)^2}{|\mathcal{K}_{j}|}},
\end{align}
where the RMSEs for $j = \mathrm{valid}$ and $j = \mathrm{test}$ are referred to as validation RMSE and test RMSE, respectively.
The validation RMSE is calculated for each training epoch.

\vspace{3pt}\noindent\textbf{FP/BP Transmission Latency.} 
The latency for transmitting FP/BP signals is calculated as follows.
Let $(P_{k})_{k\in \mathcal{K}}$ denote the measured time-varying received power values.
We use the shorthand notations $[k]$ to denote the interval $[(k - 1)\tau', k\tau']$ for $k\in\mathcal{K}$, where $\tau' = 33.3$\,ms denotes the interval between the successive received power samples.
Given that the pooling is performed with the pooling dimension of $w_{\mathrm{W}}\times w_{\mathrm{H}}$, the latency for transmitting FP signals within the interval $[k]$ is denoted by $T_{\mathrm{FP}}[k]$ and is calculated as follows:
\begin{align}
	\label{eq:forward_time}
	T_{\mathrm{FP}}[k] = \frac{L(N_{\mathrm{H}}/w_{\mathrm{H}})(N_{\mathrm{W}}/w_{\mathrm{W}})R }{W\log_{2}(1 + P_{k} / \sigma^2 W)},
\end{align}
where $\sigma^2=-173$\,dBm/Hz denotes the noise power spectral density,  $R=32$ denotes the number of bits for one pixel, and $W=40$\,MHz is the measurement bandwidth.
Likewise, the latency for transmitting BP signals within the interval $[k]$ is denoted by $T_{\mathrm{BP}}[k]$ and is calculated as follows:
\begin{align}
	\label{eq:backward_time}
	T_{\mathrm{BP}}[k] = \frac{N_{\mathrm{layer2}}R'}{W\log_{2}(1 + P_{k} / \sigma^2 W)},
\end{align}
where $N_{\mathrm{layer2}}=576$ denotes the number of weights in the upper convolutional layer, and $R'=32$ denotes the number of bits required for representing the BP gradients.
The time duration during which a stochastic gradient step is performed within the interval $[k]$ is denoted by $T_{\mathrm{step}}[k]$ and is given by $T_{\mathrm{FP}}[k] + T_{\mathrm{BP}}[k] + T_{\mathrm{comp}},$ where $T_{\mathrm{comp}}$ is the sum of time duration for calculating the FP/BP signals.

It should be noted that the training speed is affected by  $T_{\mathrm{step}}[k]$, and we also evaluate the impact of the pooling dimension on the training speed. 
For that, we calculate the time elapsed until which the $n$th stochastic gradient descent step is performed and plot its corresponding validation accuracy.
Let $N[k]\coloneqq \lfloor\tau'/T_{\mathrm{step}}[k]\rfloor$ denote the maximum number of stochastic gradient steps performed within $[k]$.
The $n$th stochastic gradient descent step is performed in a certain interval, whose index is defined as $k_n$ and is given by $\max\,\{\,k'\mid \sum_{k = 1}^{k'}N[k] \leq n\,\}$.
We calculate the time elapsed until the $n$th stochastic gradient descent step is performed, denoted by $T_n$, as follows:
\begin{align}
	T_n = \sum_{k = 1}^{k_n - 1}T_{\mathrm{step}}[k] + \!\left(n - \sum_{k = 1}^{k_n - 1} N [k] + 1\right)T_{\mathrm{step}}[{k_n}].
\end{align}

\begin{figure}[t]
	\centering
	\includegraphics[width=0.7\columnwidth]{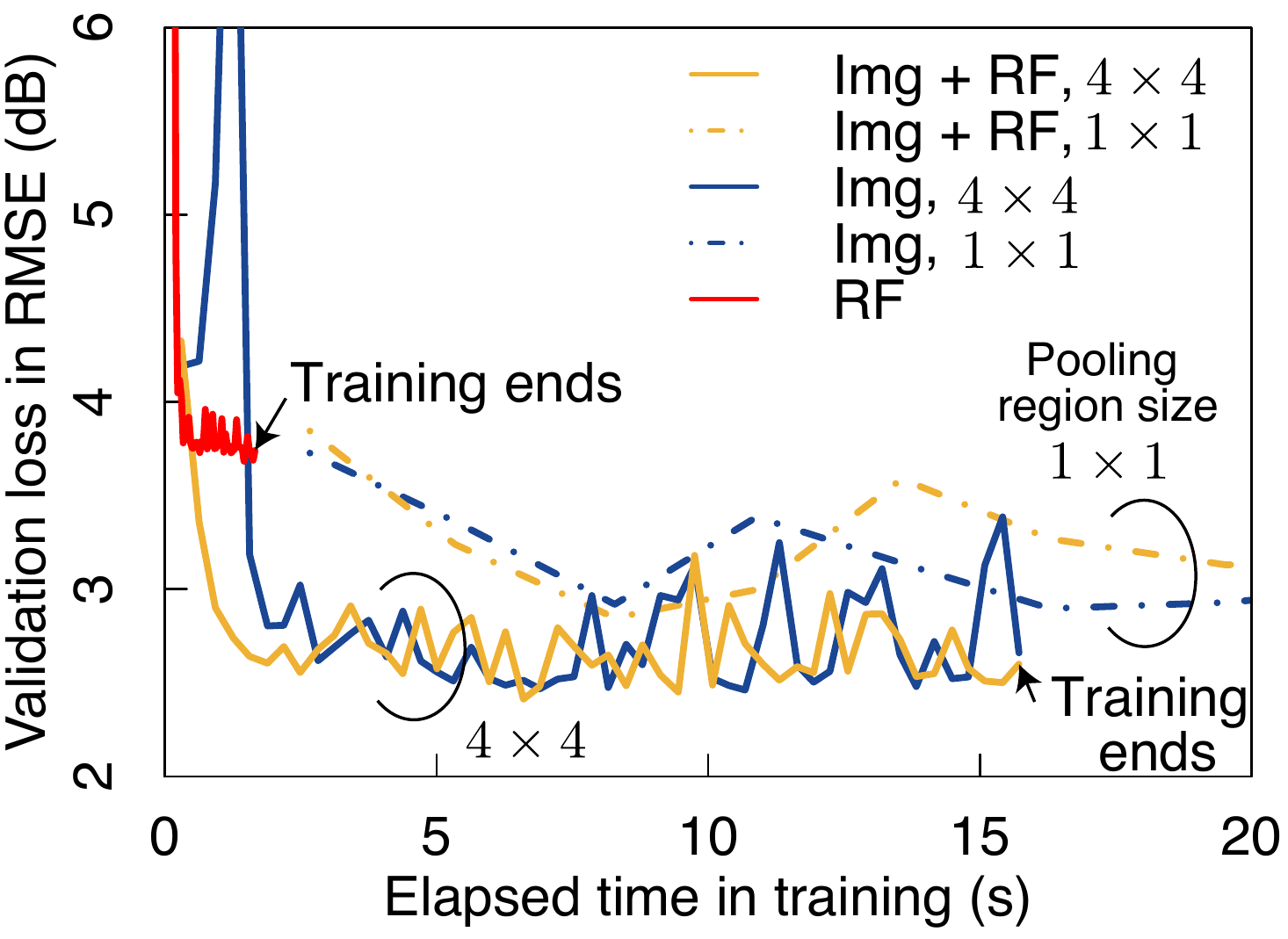}
	\caption{Impact of pooling dimension on validation accuracy in training.
	In \textsf{Img+RF} and \textsf{Img}, the larger pooling dimension results in the faster improvement on validation accuracy.}
	\label{fig:time_progress}
	\vspace{-1.8em}
\end{figure}

\vspace{3pt}\noindent\textbf{Privacy Leakage Level.} 
As the convolutional layer output becomes more similar to the input images, privacy is increasingly violated.
Thus, we quantify the privacy leakage level with the inverse of the similarity between each raw image sample $x_k$ and its CNN output.
Let $\phi(x_k)$ denote the CNN output resized such that it involves the same number of pixels as that in $x_k$ via nearest neighbor interpolation.
For measuring a similarity between $x_k$ and $\phi(x_k)$, the Euclidean-distance multidimensional scaling algorithm \cite{hout2016using} is utilized, where the low-dimensional representations of these images are embedded to an Euclidean-space, and the similarity between these images is quantitatively represented by the Euclidean-distance.
Given the measured distance $d(x_k, \phi(x_k))$, the privacy leakage is given as: $1/\max_{k\in \mathcal{K}_{\mathrm{val}}}d(x_k, \phi(x_k))$.

\begin{figure}[t]
	\centering
	\subfigure[Time series of the received power predicted 120\,ms prior to the observation when pooling dimensions correspond to $4\times 4$ and $40\times 40$.]{\includegraphics[width=0.85\columnwidth]{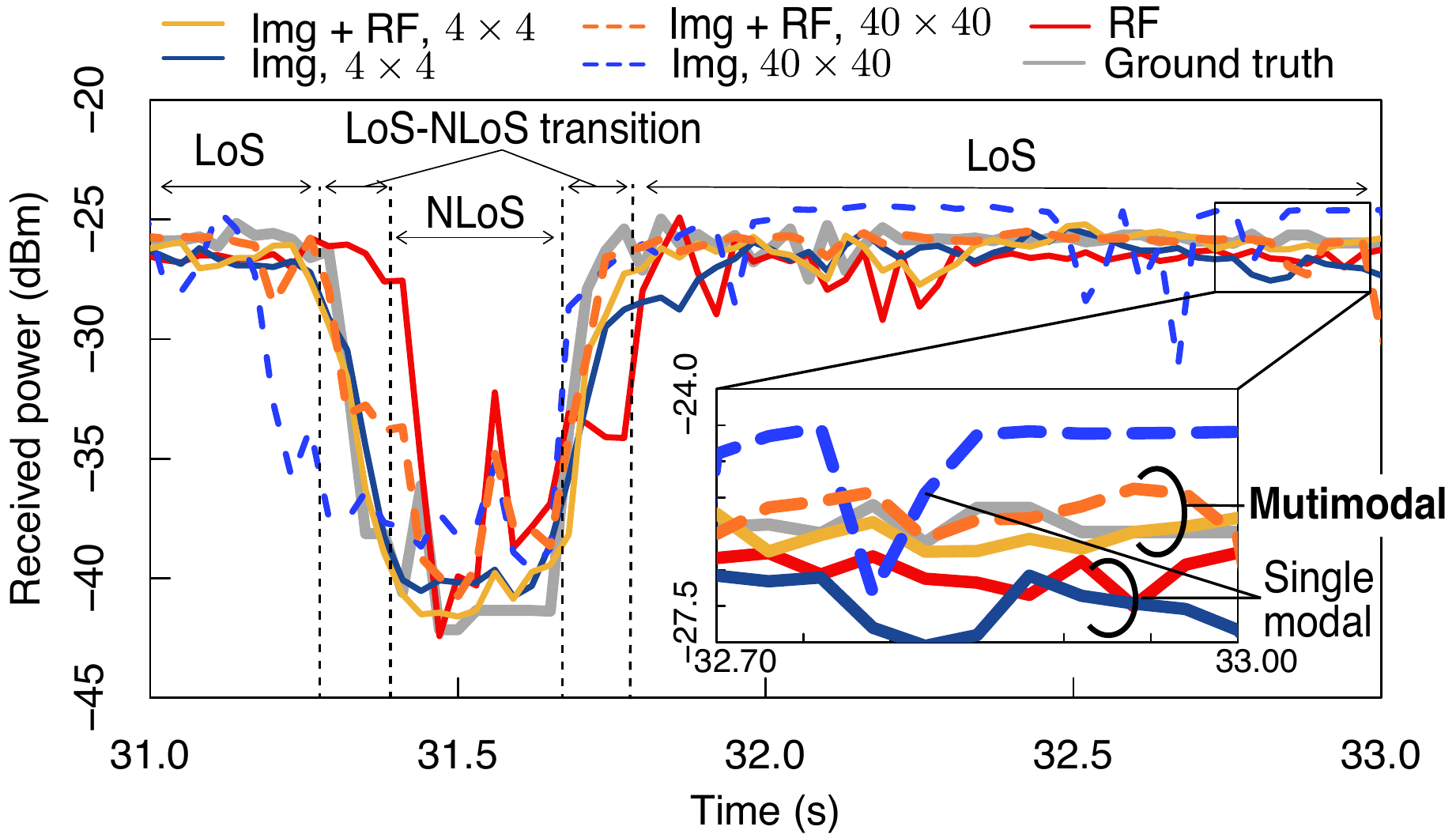}}
	\subfigure[RMSE in different channel conditions.]{\includegraphics[width=0.85\columnwidth]{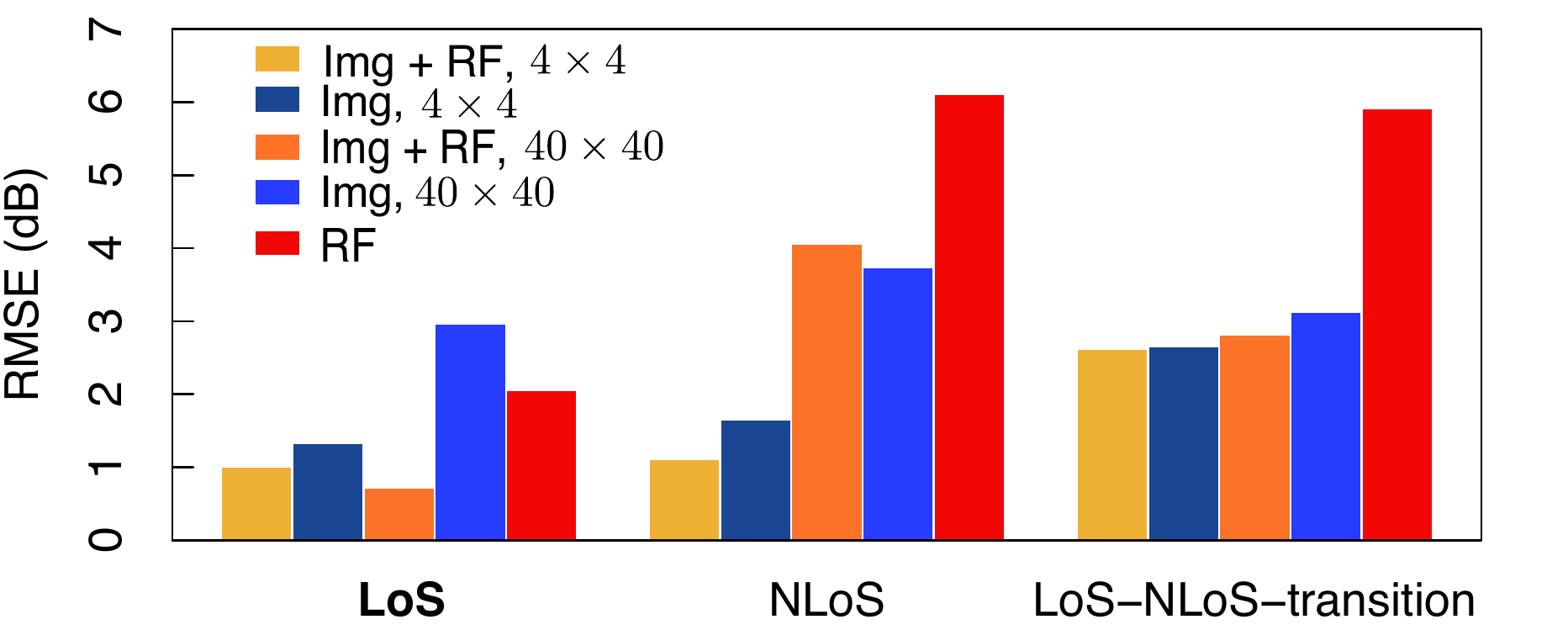}}\vspace{-0.6em}
	\caption{Received power prediction results after training.}
	\label{fig:timeseries_w4}
	\vspace{-2em}
\end{figure}

\vspace{-0.7em}

\subsection{Results and Discussions}
\noindent\textbf{Validation Accuracy in Training.} 
We analyze the impact of the pooling dimension on the validation accuracies during training.
Fig.~\ref{fig:time_progress} shows the time progress of the validation accuracy in RMSE in training.
In \textsf{Img+RF} and \textsf{Img}, the pooling dimensions of $4\times 4$ and $1\times 1$ were examined.
The computation time $T_{\mathrm{comp}}$ in either \textsf{Img+RF} and \textsf{Img} was 1.00\,ms while that in \textsf{RF} was 0.21\,ms.
The computation time $T_{\mathrm{comp}}$ in \textsf{Img+RF} and \textsf{Img} are same within the measurable accuracy (at second decimal digit); hence we treated $T_{\mathrm{comp}}$ as of the same value as 1.00\,ms in Fig.~4
\footnote{
		The same computation time within the measurable accuracy is due to a much higher computational complexity in the convolutional layers shared between \textsf{Img+RF} and \textsf{Img} compared to one in the other upper layers.
		By counting the multiplying operation in a forward propagation, we can see that the computational complexity in the convolutional layers is approximately 10--100x higher than that in the other upper layers.
		Hence, the computation in the convolutional layers is dominant, and this is the reason for the almost same computation time.
}.
The computation time is obtained by measuring the time-duration during which the aforementioned GPU computes updates of all model parameters in a stochastic gradient descent step.
In the pooling dimension of $4\times 4$, the improvement in RMSE is faster compared to the training with the pooling dimension of $1\times 1$.
This is attributed to the shorter $T_{\mathrm{FP}}[k]$ in the pooling dimension of $4\times 4$ relative to that in $1\times 1$, wherein more stochastic gradient descent steps can be performed within a certain period.
The model in \textsf{RF} can be trained 8x faster than that in either \textsf{Img+RF} and \textsf{Img} with the pooling dimension of $4\times 4$. 
This faster training is because the training in \textsf{RF} does not involve communication of FP/BP signals, i.e., $T_{\mathrm{FP}}[k] = T_{\mathrm{BP}}[k] = 0$, $\forall k\in\mathcal{K}$.
However, the RMSE reaches approximately 3.7\,dB, which corresponds to the poorer prediction performance compared to \textsf{Img+RF} and \textsf{Img} wherein the RMSE ranges from 2.7\,dB to 3.0\,dB.

\begin{figure}[t]
	\centering
	\includegraphics[width=0.75\columnwidth]{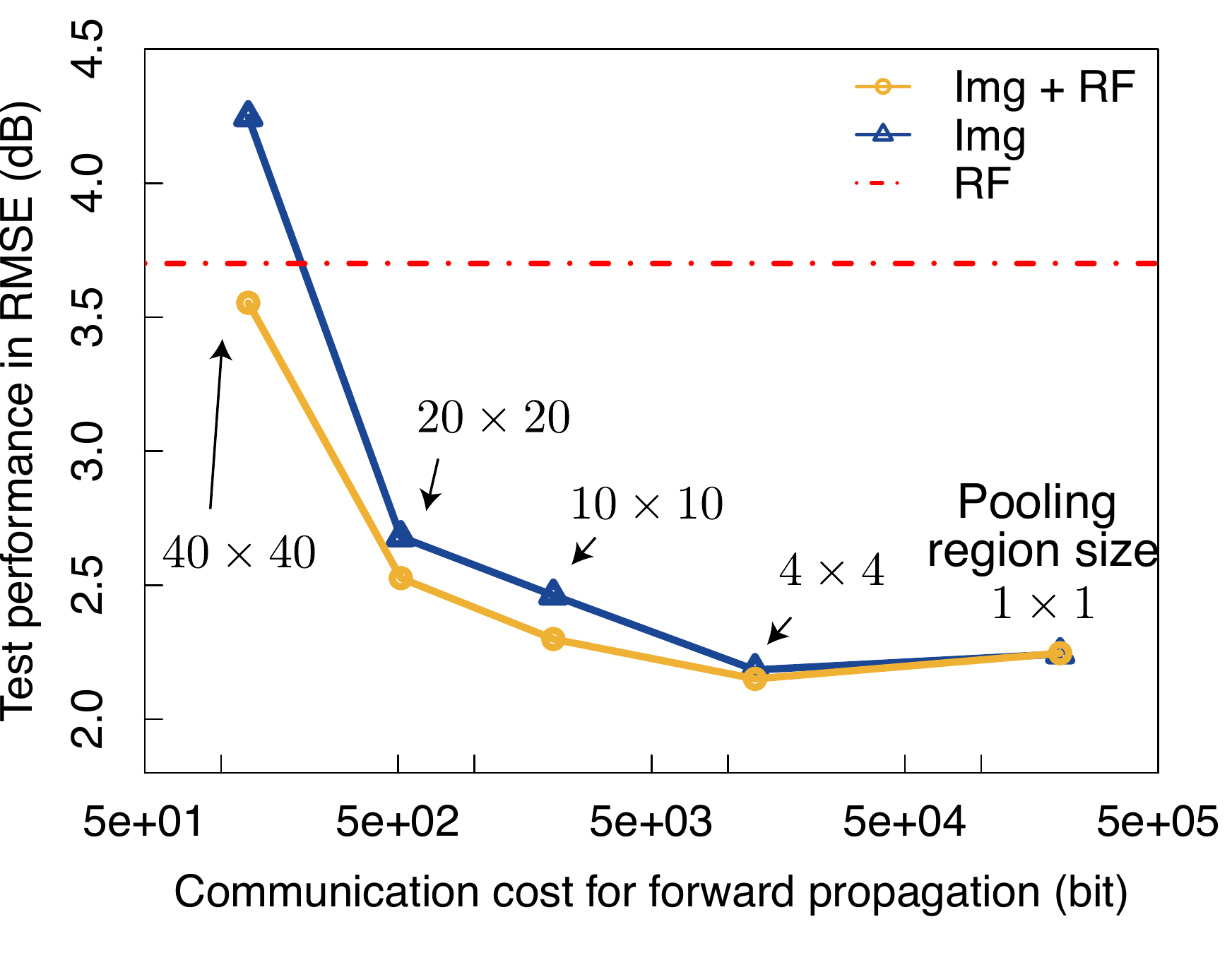}\vspace{-1em}
	\caption{Test RMSE in different pooling dimension and consequent communication cost for transmitting FP signals.}
	\label{fig:rmse_vs_cost}
	\vspace{-1em}
\end{figure}

\vspace{3pt}\noindent\textbf{Prediction after training.} 
We show that the received powers predicted by \textsf{Img + RF} match the actual received powers better than the \textsf{Img} and \textsf{RF} baselines.
Fig.~\ref{fig:timeseries_w4}(a) shows the time series of the actual received powers and of the received powers predicted 120\,ms before the actual powers were observed.
In \textsf{Img+RF} and \textsf{Img}, the models with the pooling dimensions of $4\times 4$ and $40\times 40$ are examined as an example.
First, \textsf{RF} did not match the ground truth as accurately as \textsf{Img+RF} and \textsf{Img} in particular NLoS conditions.
Focusing on LoS conditions (a zoomed-in view in Fig.~\ref{fig:timeseries_w4}(a)), \textsf{Img+RF} matches the ground truth more accurately than \textsf{Img}, which corresponds to the advantage of \textsf{Img+RF} over \textsf{Img}.
This is also quantitatively validated in Fig.~5(b) showing the RMSE in LoS, NLoS, and LoS-NLoS-transition conditions in the time-duration in Fig.~5(a), where we can see that the RMSE in \textsf{Img+RF} under a LoS condition is smaller than that in \textsf{Img}.
This can be attributed to the invariance of received powers under LoS conditions.
In LoS conditions, the input of current received powers in \textsf{Img+RF} exhibits a value similar to future received power values and can be considerably informative in terms of predicting the future received power values.
Hence, \textsf{Img+RF} predicted future received power values more accurately under LoS conditions than \textsf{Img}.

\vspace{3pt}\noindent\textbf{Test Accuracy vs. Latency and Privacy Leakage.} 
Fig.~\ref{fig:rmse_vs_cost} shows the impact of the pooling dimension on the RMSE and communication costs during inference. Drawing an inference requires a single FP transmission whose payload size is calculated by $L(N_{\mathrm{H}}/w_{\mathrm{H}})(N_{\mathrm{W}}/w_{\mathrm{W}})R$. 
The FP communication cost is monotonically decreased with the pooling dimension, and is minimized when the UE output is maximally compressed (i.e., $40\times40$ pooling dimension). By contrast, the prediction accuracy is convex-shaped over the pooling dimension. 
The maximum accuracy (i.e., minimum RMSE) is achieved when the UE output is 93\% compressed (i.e., $4\times 4$ pooling dimension), at which both accuracy and communication efficiency are improved compared to the case without compression (i.e., $1\times1$ pooling dimension). This counter-intuitive result is because the very large UE output dimension makes the LSTM layer biased only towards the image sequences while almost ignoring the received power sequences, highlighting the importance of balancing image and RF modalities.

\begin{figure}[t]
	\centering
	\includegraphics[width=0.75\columnwidth]{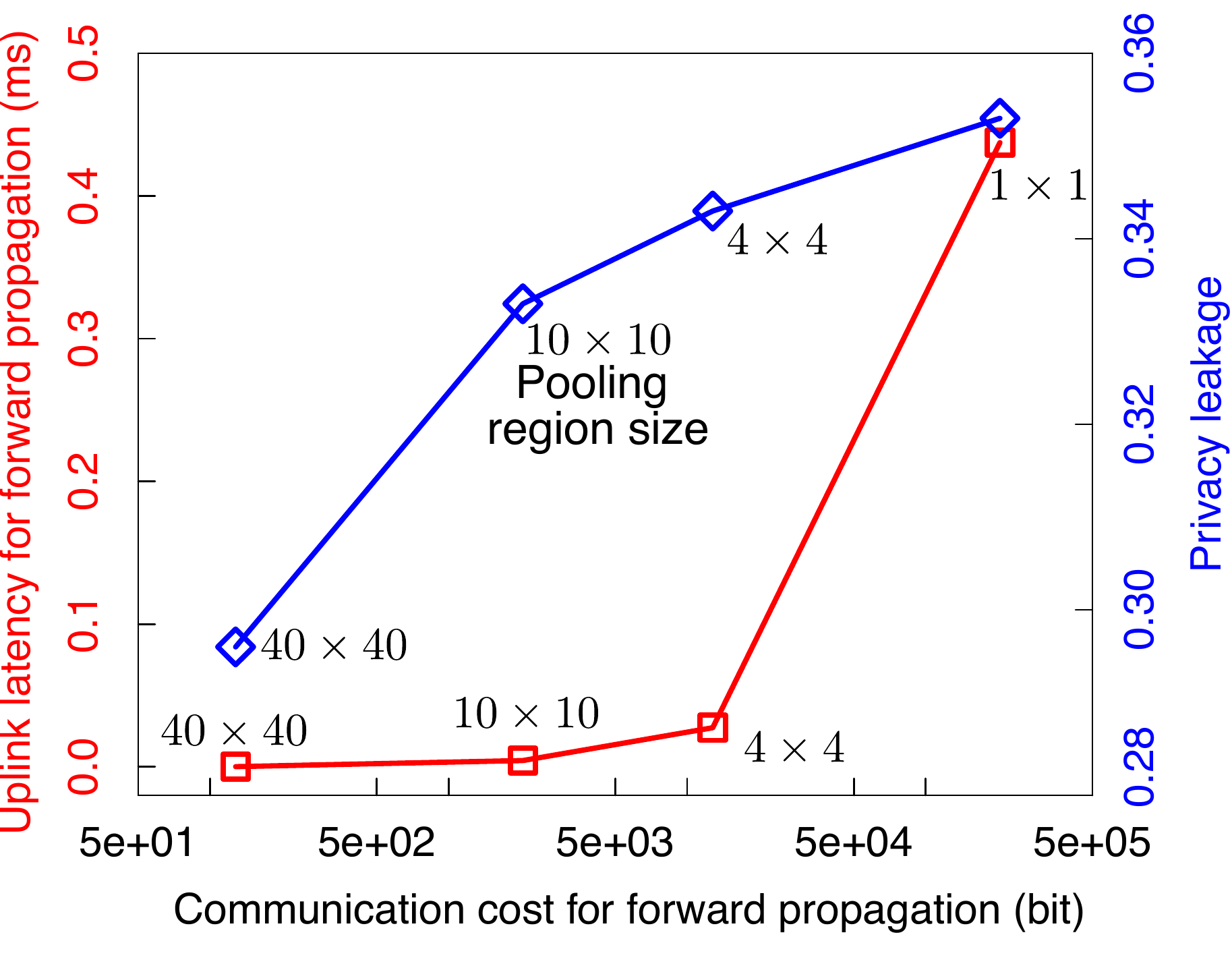}\vspace{-1em}
	\caption{Uplink latency for transmitting FP signals and privacy leakage in different pooling dimensions.}
	\label{fig:privacy_latency}
	\vspace{-1.3em}
\end{figure}

In Fig.~\ref{fig:privacy_latency}, we investigate the impact of pooling dimension on the privacy leakage level as well as the uplink latency for uploading the FP signals.
As the pooling dimension becomes larger, the raw image is 
converted to one with the smaller pixel resolutions as depicted in Fig.~\ref{fig:images}, which yields the lower latency and privacy leakage.
Specifically, in the pooling dimension of $4\times 4$, uplink latency and privacy leakage are reduced by 93\% and 2.8\%, respectively compared to the case of pooling dimension of $1\times 1$ (i.e., without compressing the convolutional layer output).

\vspace{-1em}
\section{Conclusions}
\label{sec:conclusions}
In this letter, we proposed MultSL in which a convolutional LSTM is split into two wirelessly connected segments to utilize both image and RF modalities for future mmWave received power prediction. With a single pair of image and RF signal sources, we demonstrated that optimally compressing the image segment's output dimension reduces communication payload sizes and privacy leakage without compromising the prediction accuracy. Seeking an optimal MultSL architecture for multiple image and RF signal sources under a generic network topology could be an interesting topic for future work.

\vspace{-1em}
\bibliographystyle{IEEEtran}
\bibliography{IEEEabrv,main}

\end{document}